\documentclass[prb,twocolumn,amsmath,amssymb,floatfix,eqsecnum]{revtex4}
\usepackage{graphicx}
\usepackage{bm}

\begin{document}

\preprint{cond-mat/0303001}

\title{
Quantum impurity in an antiferromagnet:\\
non-linear sigma model theory}

\author{Subir Sachdev}
\affiliation{Department
of Physics, Yale University, P.O. Box 208120, New Haven CT
06520-8120}

\author{Matthias Vojta}
\affiliation{Institut f\"ur Theorie der Kondensierten Materie,
Universit\"at Karlsruhe, Postfach 6980, D-76128 Karlsruhe,
Germany}
\date{February 28, 2003}

\begin{abstract}
We present a new formulation of the theory of an arbitrary quantum
impurity in an antiferromagnet, using the O(3) non-linear sigma
model. We obtain the low temperature expansion for the impurity
spin susceptibilities of antiferromagnets with magnetic long-range
order in the ground state. We also consider the bulk quantum phase
transition in $d=2$ to the gapped paramagnet ($d$ is the spatial
dimension): the impurity is described solely by a topological
Berry phase term which is an exactly marginal perturbation to the
critical theory. The physical properties of the quantum impurity
near criticality are obtained by an expansion in $(d-1)$.
\end{abstract}

\maketitle

\section{Introduction}
\label{sec:intro}

Recent papers \cite{sbv,vbs} have presented a general field
theoretical discussion of the low energy properties of a spin $S$
impurity embedded in an antiferromagnet or a superconductor which
is in the vicinity of a bulk spin-ordering quantum transition.
These studies were motivated by a variety of recent experiments
studying Zn and Ni impurities in the cuprate superconductors and
spin-gap compounds. The motivations and prior work have been
discussed in some detail in Ref.~\onlinecite{vbs} (hereafter
referred to as I), and so will not be repeated here. Further
theoretical \cite{sushkov,stv}, numerical
\cite{troyer,sandvik,sandvik2}, and experimental \cite{vajk} work
on these issues has also appeared, and we will discuss some of
these results below. There has also been related work on impurity
models in systems with fermionic excitations \cite{si,demler}.

The purpose of this paper is to provide additional results for the
same quantum impurity problem using a different field-theoretic
formulation. The results in I were obtained using an expansion in
$(3-d)$, where $d$ is the spatial dimensionality. Stimulated
mainly by the recent results of H\"oglund and Sandvik
\cite{sandvik}, we have succeeded in obtaining a formulation which
permits an expansion in $\epsilon=d-1$, and this will be described
in the present paper. The universal scaling structure we shall
describe below in the $(d-1)$ expansion turns out to be identical
to that obtained in I using the $(3-d)$ expansion. This is strong
evidence that a fixed point with the same scaling properties does
indeed describe the physical situation in $d=2$.

Throughout this paper, we will implicitly assume in our discussion
that $1 < d < 3$, unless stated otherwise. The only exception is
Appendix~\ref{app:d3}, where we will present results in $d=3$.
Also we will set $\epsilon=(d-1)$, whereas in I the same symbol
was used for $(3-d)$.

Let us outline the main results of I and those that will be
presented here. Consider a simple two-dimensional quantum
antiferromagnet which undergoes a quantum transition from a magnet
N\'{e}el state to a gapped, confining paramagnet with only integer
spin excitations {\em e.g.\/} a model of coupled spin ladders
\cite{ladders}. We tune the antiferromagnet across this transition
with a generalized coupling $g_0$, such that there is N\'{e}el
order for $g_0<g_c$, and a gapped paramagnet for $g_0>g_c$. Insert
an {\em arbitrary} quantum impurity ({\em e.g.\/} a vacancy) which
leads to a net deficit or excess of spin $S$ in its vicinity
(after accounting for the sublattice alternation). At a
temperature $T$ above the gapped paramagnet phase, with $g_0>g_c$
and a spin gap $\Delta$, this impurity will contribute an impurity
spin susceptibility
\begin{equation}
\chi_{\rm imp} = \frac{S(S+1)}{3T}~~~~~;~~~~~~g_0>g_c \label{e1}
\end{equation}
with exponentially small corrections as $T \rightarrow 0$ (we set
$\hbar=k_B =1$ and have absorbed factors of the gyromagnetic ratio
and the Bohr magneton in the definition of the external magnetic
field). We can view (\ref{e1}) as a definition of the value of $S$
(which must be an integer or half-odd-integer) for the quantum
impurity. In the magnetically ordered phase with $g_0<g_c$, there
are much stronger corrections to the isolated impurity behavior
because of the presence of broken spin rotation symmetry at $T=0$
and gapless excitations in the bulk; in dimensions $d \leq 2$ the
symmetry is restored at any $T>0$, and corrections to the impurity
susceptibility can be written in the scaling form\cite{vbs}
\begin{equation}
\chi_{\rm imp} = \frac{1}{T} \Phi \left( \frac{T
}{[c^{(d-2)}\rho_s]^{1/(d-1)}} \right)~~~~~;~~~~~g_0\leq g_c
\label{e2}
\end{equation}
where $\rho_s$ is the spin stiffness of the bulk ordered
antiferromagnet in the absence of the impurity, and $c$ is the
bulk spin-wave velocity. In the limit $ T \rightarrow 0$, it was
argued in I that $\Phi (0) = S^2/3$ exactly. This prediction has
been verified recently in the numerical study by H\"oglund and
Sandvik \cite{sandvik}. On the basis of the $(3-d)$ expansion, the
subleading behavior $\Phi(y \rightarrow 0) = S^2 /3 +
\mathcal{C}_3 y $, with $\mathcal{C}_3$ a universal number, was
proposed for $d=2$ in I. H\"oglund and Sandvik \cite{sandvik} also
tested this subleading behavior, and argued that it did not
hold---instead they proposed the presence of $\ln (1/T)$ term. We
will show here that their proposal is indeed correct, and that
precisely in $d=2$, the behavior in the limit $T \ll \rho_s$ is
\begin{eqnarray}
\Phi ( y \rightarrow 0) = \frac{S^2}{3} + \frac{S^2}{3 \pi}  y \ln
(1/y) + \widetilde{\mathcal{C}_3} y \nonumber\\
- \frac{S^2}{6 \pi^2} y^2 \ln (1/y) + \widetilde{\mathcal{C}}_4
y^2 + \ldots, \label{e3}
\end{eqnarray}
with $\widetilde{\mathcal{C}}_{3,4}$ unknown universal constants.
The $\ln(1/y)$ dependence is special to $d=2$ and does not appear
at any finite order in the $(3-d)$ expansion, and this is the
reason it was overlooked in I. Subleading singularities in the
small $y$ expansion do appear naturally in the $(d-1)$ expansion
presented in this paper. We also note here that (\ref{e3}) was
obtained with no assumptions on the value of $S$: the $S$
dependencies in the co-efficients are therefore exact.

The subdominant $\ln (1/T)$ dependence implied by (\ref{e3}) (and
the anomalous powers of $y$ in (\ref{eee1})) is a consequence of
spin-wave Goldstone fluctuations in $1 < d \leq 2$, and does not
involve the critical singularities at $g_0=g_c$ in an essential
way. Consequently, in $d=2$, this $\ln (1/T)$ dependence should
also be present in antiferromagnets with $g_0 \ll g_c$, which are
not especially close to any quantum critical point. In this
situation can surmise that (\ref{e3}) implies
\begin{eqnarray}
\chi_{\rm imp} &=& \frac{S^2}{3T} \Biggl[ 1  + \frac{T}{ \pi
\rho_s}\ln \left( \frac{C_1 \rho_s}{T} \right) - \frac{T^2}{2
\pi^2 \rho_s^2} \ln \left( \frac{C_2
\rho_s}{T} \right) \nonumber \\
&+& \mathcal{O}\left(\frac{T}{\rho_s}\right)^3\Biggr]~;~g_0\ll
g_c~,~T \rightarrow 0~,~d=2~~~~\label{e4}
\end{eqnarray}
where, in general, the constants $C_{1,2}$ are {\em
non}-universal; only as we approach the quantum critical point and
$\rho_s \rightarrow 0$ do $C_{1,2}$ become universal, and then
(\ref{e4}) is seen to be consistent with (\ref{e3}).  The $\ln
(1/T)$ correction in (\ref{e4}) is related to the logarithmic
frequency dependencies discussed by Nagaosa {\em et
al.}\cite{nagaosa} and Chernyshev {\em et al.} \cite{castro}. To
the extent that sharp spin-waves are also present in ordered
metallic antiferromagnets, (\ref{e4}) may also apply to such
systems \cite{morr}.

As we will see shortly, (\ref{e4}) is obtained for the case where
the coupling between the impurity and the bulk antiferromagnet has
scaled to infinity. This implies that at low energies the impurity
moment is effectively locked along the direction of the local
orientation of the bulk antiferromagnetic order. While such
locking is appropriate near the quantum critical point, it is not
a priori clear whether it should also hold at low $T$ above a well
ordered antiferromagnet with $g_0 \ll g_c$. We will briefly
address this issue by also examining the case of finite coupling
(see Appendix~\ref{app:gamma}): we find that the co-efficient of
the $(1/\rho_s) \ln (1/T)$ term in (\ref{e4}) remains universal,
but there are non-universal corrections to the $T \ln (1/T)$ term.

A separate category of our results concern $\chi_{\rm imp}$ at the
quantum critical point, $g_0=g_c$. These correspond to the large
$y$, $T \gg \rho_s$, limit of (\ref{e2}). Here, it was argued in I
that
\begin{equation}
\Phi ( y \rightarrow \infty ) = \mathcal{C}_1 \label{e5}
\end{equation}
with $\mathcal{C}_1$ a universal number. A $(3-d)$ expansion for
$\mathcal{C}_1$ was provided in I, and it contained non-trivial
corrections to the free moment value of $S(S+1)/3$. Sushkov
\cite{sushkov} has questioned the existence of such corrections,
but we reply to his arguments in Appendix~\ref{app:sushkov}. The
present paper will show that (\ref{e5}) is obeyed also in the
$(d-1)$ expansion: in this case the $(d-1)$ expansion provides
terms as corrections to the `classical' moment value of $S^2 /3$,
and details of this appear in the body of the text, and the final
result is in (\ref{r12}).

A number of other results for universal properties of the impurity
correlations were provided in I using the $(3-d)$ expansion. All
of these can also be computed in the $(d-1)$ expansion, and in
every case we find complete agreement in the structure of the
scaling properties. Details of such computations also appear in
the body of this paper.

The following section will introduce the non-linear sigma model
field theory which describes the dynamics of an impurity in a
quantum antiferromagnet. Section~\ref{sec:perturb} will then
discuss the perturbative structure of this theory, with details of
the perturbative computations appearing in
Appendix~\ref{app:perturb}. We will show how to deduce low
temperature properties using this perturbation theory. Finally,
Section~\ref{sec:rg} presents a renormalization analysis which
allows us to deduce the physical characteristics of the critical
point.

\section{Field Theory}
\label{sec:ft}

This section will introduce the field-theoretical formulation of
the quantum impurity dynamics which enables an expansion of its
universal properties in the $(d-1)$ expansion. In contrast to our
earlier $(3-d)$ expansion, which used a `soft-spin' formulation of
the bulk antiferromagnetic fluctuations, the present $(d-1)$
expansion will use the `fixed-length' representation  of the O(3)
non-linear sigma model.

We begin by recalling our earlier `soft-spin' formulation. The
bulk spin fluctuations of the antiferromagnet are represented by
the real field $\phi_{\alpha} (x, \tau)$, with $\alpha=1 \ldots 3$
an index representing the spin component, $x$ a $d$-dimensional
spatial co-ordinate, and $\tau$ is imaginary time. The impurity
spin is placed at the origin of co-ordinates $x=0$, and is
represented by a unit length field $n_{\alpha} (\tau)$, and the
bulk and impurity fluctuations are coupled in the partition
function
\begin{eqnarray}
&&\widetilde{\mathcal{Z}} = \int \mathcal{D} \phi_{\alpha} (x,
\tau) \mathcal{ D} n_{\alpha}
(\tau ) \delta \left( n_{\alpha}^2 - 1 \right) \nonumber \\
&&~~~~~~~~~~~~~~~~~~~~~~~~ \exp\left(-
\widetilde{\mathcal{S}}_{\text{b}}
[\phi_{\alpha}] - \widetilde{\mathcal{S}}_{\text{imp}} \right) \nonumber \\
&& \widetilde{\mathcal{S}}_{\text{imp}} =  \int_0^{1/T} d \tau
\bigg[ i S
A_{\alpha} (n) \frac{d n_{\alpha}(\tau)}{d \tau} \nonumber \\
&&~~~~~~~~~~~~~~~~~~~~~~~~~ - \gamma S \phi_{\alpha} (x = 0, \tau)
n_{\alpha} (\tau) \bigg]. \label{simp}
\end{eqnarray}
The transition in the bulk antiferromagnet is described by the
usual $\phi_{\alpha}^4$ theory which is represented by
$\widetilde{\mathcal{S}}_b [ \phi_{\alpha}]$ as in I. The first
term in the impurity action $\widetilde{\mathcal{S}}_{\text{imp}}$
is the Berry phase of the impurity at site $r$: and $A_{\alpha}
(n)$ is a `Dirac monopole' function which satisfies
\begin{equation}
\epsilon_{\alpha\beta\gamma} \frac{\partial A_{\gamma}
(n)}{\partial n_{\beta}} = n_{\alpha}. \label{diracmono}
\end{equation}
Finally, $\gamma$ is the coupling between the impurity and bulk
degrees of freedom which will be important in our considerations
below. At the $\gamma=0$ fixed point, the bulk and boundary
degrees of freedom are decoupled, and the coupling $\gamma$ is a
relevant perturbation with scaling dimension $(3-d-\eta)/2$
($\eta$ is  the anomalous dimension of the bulk critical point,
and its value is very close to zero). The small scaling dimension
of $\gamma$ near $d=3$ was the key feature which was used to
generate the $(3-d)$ expansion of the coupled bulk-impurity
theory.

Let us now turn to spatial dimensions just above $d=1$. For the
bulk theory, it is known that an expansion of the critical
properties can be generated in a $\epsilon=d-1$ expansion by
representing the bulk spin fluctuations by a fixed-length field
$N_{\alpha} (x, \tau) \propto \phi_{\alpha} (x, \tau)$, and with
the action of the O(3) non-linear sigma model \cite{bz}. At the
same time, the coupling $\gamma$ has a scaling dimension $\approx
1$, and so is strongly relevant near $\gamma=0$. This suggests
that a better approach now would be to start near the
$\gamma=\infty$ limit. At $\gamma=\infty$, the impurity degrees of
freedom $n_{\alpha} (\tau)$ would follow the bulk spin
fluctuations perfectly, and hence $n_{\alpha} (\tau) = N_{\alpha}
(x=0, \tau)$. In this manner we obtain the central field theory of
interest in this paper
\begin{eqnarray}
&& \mathcal{Z} = \int \mathcal{ D} N_{\alpha} (x, \tau) \delta
\left( N_{\alpha}^2 - 1 \right) \exp\left(- \mathcal{S}_{\text{b}}
[N_{\alpha}] - \mathcal{S}_{\text{imp}} \right) \nonumber \\
&& \mathcal{S}_{\text{b}}[N_{\alpha}] = \frac{1}{2cg_0} \int d^d x
\int_0^{1/T} d \tau \left[ \left( \partial_{\tau} N_{\alpha}
\right)^2 + c^2 \left( \nabla_x N_{\alpha} \right)^2 \right]
\nonumber
\\ && \mathcal{S}_{\text{imp}} =  \int_0^{1/T} d \tau \bigg[ i S
A_{\alpha} (n) \frac{d n_{\alpha}(\tau)}{d \tau}\bigg]\nonumber
\\&&~~~~~~~~~~~~~~~~\mbox{with $n_{\alpha} (\tau) \equiv N_{\alpha} (x=0,\tau)$}
\label{simpn}
\end{eqnarray}
We will set $c=1$ in the remainder of the paper as it does not
appear in any essential manner in any of our expressions, and it
can be easily re-inserted by dimensional analysis. The Berry phase
in $\mathcal{S}_{\text{imp}}$ is invariant under global spin
rotations and is independent of the gauge choice for $A_{\alpha}$.
Using an analysis very similar to that presented in I, it can be
shown, order by order in $(d-1)$, that there are no relevant
perturbations to the terms shown in (\ref{simpn}) at the quantum
critical point. Furthermore, the Berry phase
$\mathcal{S}_{\text{imp}}$ turns out to be an {\em exactly
marginal} perturbation to the bulk critical point, whose coupling
constant ($S$) is protected by its topological nature. There is
only a single remaining coupling constant in $\mathcal{Z}$, and
that is the bulk coupling $g_0$, and its renormalization is
unaffected by the presence of a single impurity spin. As in
Ref.~\onlinecite{bz}, all bulk and impurity spin correlations can
be computed order by order in $g_0$ in a diagrammatic perturbation
theory. We defer discussion of the structure of this diagrammatic
expansion to Appendix~\ref{app:perturb}. We note here that this
perturbation theory makes no assumptions on the value of the
impurity spin $S$, and the Berry phase is fully accounted for at
each order in the perturbation theory in $g_0$.

It is worth noting here that a perturbation theory in powers of
$g_0$ can also be generated for an arbitrary value of $\gamma$,
with $n_{\alpha} (\tau) \neq N_{\alpha} (x=0, \tau)$ (no expansion
in $\gamma$ or $S$ is needed here). In this case there is an
additional gapped excitation corresponding to the deviation of the
impurity spin from the bulk antiferromagnetic spin fluctuations
(the gap of this excitation is of order $\gamma$). This
perturbation theory is somewhat more cumbersome and is discussed
briefly in Appendix~\ref{app:gamma}.

\section{Perturbation theory at low $T$}
\label{sec:perturb}

Before embarking upon the subtleties of a renormalization group
analysis (and the associated analytic continuation in
dimensionality), it is useful to examine the expressions in
Appendix~\ref{app:impt} directly in $1< d \leq 2$, in a regime
where perturbation theory is valid. Perturbation theory holds for
small $g_0$, or alternatively for `large' $\rho_s$. Consequently,
direct perturbative results can be obtained in the
renormalized-classical region with $T \ll \rho_s$.

We discuss some important features of the perturbation theory
here, with further details appearing in
Appendix~\ref{app:perturb}. For dimensions $1 < d \leq 2$, there
is long-range magnetic order for $g_0< g_c$ at $T=0$, but rotation
symmetry is restored at any $T>0$. This singular phenomenon
accounted for by a two-step integration procedure which has been
discussed in detail in Sections 6.3.2 and 7.1.2 of
Ref.~\onlinecite{book}: first we integrate out the modes with
Matsubara frequency $\omega_n \neq 0$, and then subsequently
perform a rotational average over the static modes by an exact
procedure. The first step is easily performed by a perturbation
theory in which we assume that the local magnetic order is
polarized along, say, the $(0,0,1)$ direction. We obtain an
expansion for the free energy in the presence of an applied
magnetic field $H_{\alpha}$, which we assume has the value
\begin{equation}
H_{\alpha} = (H_{\perp}, 0, H_{\parallel}). \label{a6}
\end{equation}
This expansion is discussed in some detail in
Appendix~\ref{app:impt}, and yields the following expression for
the free energy
\begin{eqnarray}
\mathcal{F} &=& - T \ln \mathcal{Z} \nonumber \\
&=& \mathcal{F}_0 - m H_{\parallel} - \frac{1}{2} \chi_{\parallel}
H_{\parallel}^2 - \frac{1}{2} \chi_{\perp} H_{\perp}^2 ;\label{a7}
\end{eqnarray}
here $\mathcal{F}_0$ is the free energy in zero field. In
(\ref{a7}) $m$ has the apparent interpretation of the local
magnetic moment of the impurity, while $\chi_{\perp,\parallel}$
appear to be the transverse and longitudinal susceptibilities.
However, it must be kept in mind that we are working in a $T>0$
regime where the magnetic order is ultimately averaged over and so
$m$, $\chi_{\parallel,\perp}$ are merely intermediate quantities
which arise in our computation, and do not have independent
physical meaning. For $g_0 < g_c$ the moment $m$ is quantized
exactly at the value $m=S$ at $T=0$, but corrections do appear at
$T>0$, as shown in Appendix~\ref{app:impt}. Following the method
discussed in Section 6.3.2 of Ref.~\onlinecite{book}, to the order
in perturbation theory being considered here, the second step of
rotational averaging over the directions of the local
magnetization leads to the following expression for the physical
magnetic susceptibility
\begin{equation}
\chi = \frac{m^2}{3T} + \frac{1}{3} \chi_{\parallel} +
\frac{2}{3}\chi_{\perp}. \label{a8}
\end{equation}
Only the final quantity $\chi_{\text{imp}}$ is a physical
observable at $T>0$.

We can divide the contributions to the quantities in (\ref{a8}) to
those arising from the bulk antiferromagnet (which are
proportional to its volume) and to those associated with the
impurity. First, for completeness, we recall results for the bulk
susceptibilities, which are implicitly expressed per unit volume;
there is no bulk contribution to the magnetic moment $m$. The
results of bare perturbation theory for the bulk susceptibilities
quantities are given in (\ref{perpb}) and (\ref{parb}). We
re-express the results by replacing $g_0$ by the physical
$\rho_s$; these two quantities are related by \cite{book}
\begin{equation}
\rho_s = \frac{1}{g_0} \left[ 1 - g_0 \int \frac{d^d k}{(2 \pi)^d}
\frac{1}{2 k} + \mathcal{O} (g_0^2) \right]. \label{rhos}
\end{equation}
In this manner, we obtain
\begin{eqnarray}
\chi_{\perp,\text{b}} &=& \rho_s - \int \frac{d^d k}{(2 \pi)^d}
\left(
\frac{1}{k (e^{k/T} - 1)} - \frac{T}{k^2}\right) \nonumber \\
\chi_{\parallel,\text{b}} &=&  \int \frac{d^d k}{(2 \pi)^d} \left(
\frac{1}{2 \sinh^2 (k/(2T))} - \frac{2T}{k^2}\right).
\end{eqnarray}
Notice that both expressions have an ultraviolet divergence for $d
\geq 2$, and so depend on the upper cutoff of the momentum
integration. However, this divergence disappears in the physical
bulk susceptibility
\begin{eqnarray}
\chi_{\text{b}} &=& \frac{1}{3} \chi_{\parallel,\text{b}} +
\frac{2}{3}\chi_{\perp,\text{b}} \nonumber \\
&=& \frac{2}{3}\rho_s + \frac{1}{3}\int \frac{d^d k}{(2 \pi)^d}
\left(\frac{1}{2 \sinh^2 (k/(2T))} \right. \nonumber \\
&~&~~~~~~~~~~~~~~~~~~~~\left. - \frac{2}{k (e^{k/T} - 1)} \right)
\nonumber \\
&=& \frac{2\rho_s}{3 c^2} + \frac{T}{3 \pi c^2}~~\mbox{for $d=2$}.
\label{chib}
\end{eqnarray}
The last expression has been evaluated in $d=2$, and we have
re-inserted factors of $c$; this result has appeared earlier in
the literature \cite{hasen,csy}.

It is also interesting to see how (\ref{chib}) can also be
obtained by the dimensionally regularized expressions in
(\ref{perpb}) and (\ref{parb}).  In dimensional regularization,
the relationship (\ref{rhos}) becomes simply $\rho_s = 1/g_0$;
substituting this into the integrals already evaluated in
(\ref{perpb}) and (\ref{parb}) we obtain
\begin{equation}
\chi_{\text{b}}=\frac{2\rho_s }{3} - \frac{2}{3} \pi^{d/2-2}
\Gamma(2-d/2) \zeta (2-d) T^{d-1}; \label{chib1}
\end{equation}
here $\Gamma (s)$ is the Gamma function, and $\zeta (s)$ is the
Riemann zeta function. Eqn. (\ref{chib1}) agrees with (\ref{chib})
after using $\zeta(0)=-1/2$.

After subtracting out the bulk contributions to (\ref{a7}), we are
left with the impurity magnetization and susceptibilities. These
can be computed by the same method as for the bulk
susceptibilities. We will discuss  the impurity response to a
uniform magnetic field at $T>0$ in Section~\ref{sec:imp}.
Section~\ref{sec:loc} will consider the case of a local magnetic
field applied only in the vicinity of the impurity site, while
Section~\ref{sec:zero} generalizes our results to a uniform
magnetic field at $T=0$.

\subsection{Impurity susceptibility at $T>0$}
\label{sec:imp}

Evaluating first the frequency summations in (\ref{mexp}), then
inserting (\ref{rhos}), we obtain the following result for the
impurity magnetic moment, $m$, as an expansion in $1/\rho_s$:
\begin{eqnarray}
m &=& S \left[ 1 + \frac{T}{\rho_s} \left\{\int \frac{d^d k}{(2
\pi)^d} \left(\frac{1}{k^2} - \frac{1}{4 T^2 \sinh^2 (k/(2T))}
\right) \right\}\right. \nonumber \\ &~&~~~~\times \left\{1 +
\frac{T}{\rho_s} \int \frac{d^d p}{(2 \pi)^d} \left(\frac{1}{pT
(e^{p/T}-1)} \right.\right. \nonumber \\
&~&~~~~~~~~~~~~~~~~~~~\left.\left.\left.- \frac{1}{4 T^2 \sinh^2
(p/(2T))} \right)\right\}\right] \label{be1}
\end{eqnarray}
Notice that while intermediate terms (like those in (\ref{rhos}))
had a linear ultraviolet divergence in $d=2$, the final
expressions only have a logarithmic dependence upon the
ultraviolet cutoff in $d=2$. Unlike the case for the bulk
susceptibility above, this divergence will not cancel against any
other term. We can evaluate the integrals in (\ref{be1}) with a
cutoff $\Lambda$ in $d=2$ and obtain in the limit $\Lambda/T
\rightarrow \infty$
\begin{equation}
m = S \left[ 1 + \left( \frac{T}{2 \pi \rho_s} - \frac{T^2}{4
\pi^2 \rho_s^2} \right) \ln \frac{\Lambda}{T} \right]~~;~~d=2
\label{be2}
\end{equation}

As we noticed above for the bulk susceptibility, the result
(\ref{be2}) can also be obtained in a somewhat simpler manner by
the dimensionally regularized expressions in
Appendix~\ref{app:impt}. Using the integrals already evaluated in
(\ref{mexp}) we obtain
\begin{equation}
m = S  + m_1 \frac{T^{d-1}}{\rho_s} + m_2
\frac{T^{2(d-1)}}{\rho_s^2}  \label{be3}
\end{equation}
where
\begin{eqnarray}
m_1 &=& \frac{S(1-d)}{2 \pi^{2-d/2}} \Gamma(1-d/2) \zeta(2-d)
\nonumber \\
m_2 &=& \frac{S(d^2-3d+2)}{4 \pi^{4-d}} [\Gamma(1-d/2)
\zeta(2-d)]^2 .~~~~~\label{ae4}
\end{eqnarray}
Finally, we take the limit $d \rightarrow 2$ in (\ref{ae4}). This
is found to be singular, as $m_{1,2}$ both develop poles in
$(2-d)$. In particular $m_1 \rightarrow 1/(2 \pi (2-d))$ and $m_2
\rightarrow -1/(4 \pi^2 (2-d))$. As is conventional, we may
identify the poles in $(2-d)$ with the logarithmic dependence upon
the cutoff, and in this manner our earlier result (\ref{be2}) is
seen to be perfectly consistent with (\ref{be3}) and (\ref{ae4}).

We may proceed in a similar manner to an evaluation of the
expressions for $\chi_{\perp,\text{imp}}$ and
$\chi_{\parallel,\text{imp}}$ in (\ref{perpexp}) and
(\ref{parexp}). Here rather than using a momentum cutoff, we use
the insights gained above to proceed with the simpler dimensional
regularization method. Inserting the resulting expressions into
(\ref{a8}) we obtain the final result
\begin{equation}
\chi_{\text{imp}}= \frac{1}{T} \left[ \frac{S^2}{3} + \chi_1
\frac{T^{d-1}}{\rho_s} + \chi_2 \frac{T^{2(d-1)}}{\rho_s^2}
\right] \label{ae5}
\end{equation}
where
\begin{eqnarray}
\chi_1 &=& \frac{S^2 (1-d)}{3 \pi^{2-d/2}} \Gamma(1-d/2)
\zeta(2-d)
\nonumber \\
\chi_2 &=& \frac{S^2}{24 \pi^{4-d}} [\Gamma(1-d/2)]^2 \bigl[ (2d^2
-6d + 5) \zeta(4-2d) \nonumber \\ &~&~~~+ 2 (3d^2 - 8d + 5)
(\zeta(2-d))^2 \bigr] \label{ae6}
\end{eqnarray}
As below (\ref{ae4}), upon taking the limit $d \rightarrow 2$, the
expressions in (\ref{ae6}) are seen to have simple poles in
$(2-d)$ with $\chi_1 = S^2 /(3 \pi (2-d))$ and $\chi_2 = -S^2/(6
\pi^2 (2-d))$;  we replace the poles by $\ln(\Lambda/T)$ and
thence obtain the result (\ref{e4}) announced in the introduction.
We have also checked (\ref{e4}) directly in $d=2$, by estimating
the value of momentum integrals with a finite cutoff, using
integrands similar to (\ref{be1}) obtained from
Appendix~\ref{app:impt}.

Close to the critical point, the expansion (\ref{ae5}) implies
that for general $1 < d < 2$, the small $y$ expansion of the
scaling function $\Phi (y)$ in (\ref{e2}) has the structure
\begin{equation}
\Phi (y \rightarrow 0) = \frac{S^2}{3} + \chi_1 y^{d-1} + \chi_2
y^{2(d-1)} + \ldots \label{eee1}
\end{equation}
with the universal numbers $\chi_{1,2}$ specified in (\ref{ae6}).
The $d \rightarrow 2$ limit of $\chi_{1,2}$ then leads directly to
(\ref{e3}). We are unable to obtain the values of the universal
constants $\widetilde{C}_{3,4}$ because accurate results for the
critical point are only possible for $d$ close to 1, but in that
case the logarithms of (\ref{e3}) are absent.

A significant feature of these results for $\chi_{\text{imp}}$ is
that while there is a $(T/\rho_s) \ln (1/T)$ term, the
$(T/\rho_s)^2 \ln^2 (1/T)$ terms have cancelled against each
other; alternatively stated, the double pole in $(2-d)$ that is
apparently present in (\ref{ae6}) (associated with
$[\Gamma(1-d/2)]^2$) turns out to have vanishing residue because
$\zeta (0) = -1/2$. This is an indication that this particular log
singularity does not exponentiate upon inclusion of higher order
terms, and is rather a consequence of Goldstone spin-wave
fluctuations, as opposed to a critical singularity.

\subsection{Local susceptibility at $T>0$}
\label{sec:loc}

We now consider the {\em local\/} susceptiblity,
$\chi_{\text{loc}}$ which is the response to a field applied at
the impurity site only. This is to be distinguished from the
impurity susceptibility which is the response to a uniform field,
after subtracting out the bulk contribution. The small $g_0$
expansion for $\chi_{\text{loc}}$ is discussed in
Appendix~\ref{app:loc}. The relationship (\ref{a8}) now
generalizes to
\begin{equation}
\chi_{\text{loc}} = \frac{m_{\text{loc}}^2}{3T} + \frac{1}{3}
\chi_{\parallel, \text{loc}} + \frac{2}{3}\chi_{\perp,\text{loc}},
\label{loc1}
\end{equation}
and expressions for the terms on the r.h.s. appear in
(\ref{apploc1}). Evaluating the frequency summations and the
momentum integrals with a cutoff $\Lambda$ we obtain in $d=2$
\begin{equation}
\chi_{\text{loc}} = \frac{S_{\text{loc}}^2}{3T} +
\left(\frac{S^2}{3 \pi \rho_s} + \frac{C_3 \Lambda+C_4
T}{\rho_s^2} \right)\ln \left( \frac{\Lambda}{T} \right) + \ldots
\label{loc2}
\end{equation}
Here $S_{\text{loc}}$ is a non-universal impurity moment which
depends upon microscopic details like the local coupling constants
and the precise location over which the field is applied. It is,
in general, not equal to $S$, the moment which appears in
$\chi_{\text{imp}}$. Similarly, $C_{3,4}$ are  non-universal
numbers. Note however, that as in (\ref{e4}), there is no term of
order $(T/\rho_s^2) \ln^2 (1/T)$.

\subsection{Zero temperature response to an applied field}
\label{sec:zero}

The divergent impurity susceptibilities obtained above as
$T\rightarrow 0$ suggest that the response to a field will be
singular at $T=0$.

At $T=0$, the magnetic symmetry is broken for $d >1$ and small
$g_0$, and so the quantities $m$, $\chi_{\parallel,\text{imp}}$,
and $\chi_{\perp,\text{imp}}$ retain their separate physical
identities and can be distinguished experimentally.

The calculation of the impurity response to a magnetic field at
$T=0$ proceeds in a manner similar to that at $T>0$. The first
crucial observation is that we now have\cite{vbs}
\begin{eqnarray}
m&=&S \nonumber \\
\chi_{\parallel,\text{imp}} &=& 0 \label{x1}
\end{eqnarray}
to all orders in $g_0$. This is a consequence of a `gauge
invariance' of the action $\mathcal{Z}$ associated with the
preserved symmetry of rotations about the $z$ axis, and the
transformation (\ref{a1}). Explicitly, it is not difficult to
check that upon converting the frequency summations to integrals
in (\ref{mexp}) and (\ref{parexp}), and evaluating the frequency
integrals, the results in (\ref{x1}) hold to order $g_0^2$---this
is a strong check on our computations.

It remains to compute the transverse susceptibility
$\chi_{\perp,\text{imp}}$. Because of the broken spin rotation
symmetry, this quantity is not protected by gauge invariance (the
gauge symmetry is `broken'), and it has non-zero contributions at
each order in perturbation theory. However, certain terms in the
perturbation theory have an infrared divergence for $d \leq 2$ in
the presence of fully O(3) symmetric Hamiltonian, and so we
examine the full non-linear dependence of the impurity free energy
on the applied field, $\mathcal{F}_{\text{imp}} (H_{\perp})$. The
perturbative computation of $\mathcal{F}_{\text{imp}}$ is
described in Appendix~\ref{app:zero}, and from (\ref{zero2}) we
obtain for $1< d < 2$
\begin{eqnarray}
\mathcal{F}_{\text{imp}} (H_{\perp}) - \mathcal{F}_{\text{imp}}
(0) &=& - f_1 \frac{H_{\perp}^d}{\rho_s}  + \ldots \nonumber \\
f_1 &\equiv & \frac{S^2 \Gamma(1-d/2)}{2 ( 4 \pi)^{d/2}}.
\label{x2}
\end{eqnarray}
This result, and the structure of the perturbation theory in
Appendix~\ref{app:zero}, suggest the follow universal scaling form
for the critical behavior of $\mathcal{F}_{\text{imp}}
(H_{\perp})$ near the critical point:
\begin{equation}
\mathcal{F}_{\text{imp}} (H_{\perp}) - \mathcal{F}_{\text{imp}}
(0) = -H_{\perp} \Phi_{\mathcal{F}} \left(
\frac{H_{\perp}}{\rho_s^{1/(d-1)}} \right); \label{x3}
\end{equation}
the results here and in I imply that this scaling form holds for
all $1 < d < 3$. The results of I implicitly assumed an analytic
dependence of $\mathcal{F}_{\text{imp}}$ on small $H_{\perp}^2$,
so that $\Phi_{\mathcal{F}} (y) \sim y$ for small $y$. However,
our computations here show that this analyticity holds only for
$d>2$, and that there is a leading non-analytic dependence with
$\Phi_{\mathcal{F}} (y\rightarrow 0) \sim y^{d-1}$ for $1<d<2$.
Precisely in $d=2$, there is a pole in $f_1$ defined in
(\ref{x2}), and as in our discussion for $m$, this implies a
logarithmic singularity:
\begin{equation}
\Phi_{\mathcal{F}} (y \rightarrow 0) = \frac{S^2}{4 \pi} y \ln
(1/y) + \widetilde{\mathcal{C}}_{\mathcal{F}}y + \ldots,
\label{x4}
\end{equation}
where $\widetilde{\mathcal{C}}_{\mathcal{F}}$ is an unknown
universal number. In the ordered state in $d=2$, well away from
the critical point, we have from (\ref{zero2}), and as in
(\ref{e4})
\begin{eqnarray}
\mathcal{F}_{\text{imp}} (H_{\perp}) - \mathcal{F}_{\text{imp}}
(0) &=& -\frac{S^2 H_{\perp}^2}{4 \pi \rho_s} \ln \left( \frac{C_3
\rho_s}{H_{\perp}} \right)\nonumber \\
&~&;~~~g_0\ll g_c~,~H_{\perp} \rightarrow 0,~~~~~~\label{x5}
\end{eqnarray}
where $C_3$ is a non-universal number. The logarithmic
singularities in (\ref{x4}) and (\ref{x5}) can be cut-off by
spin-anisotropies in the underlying Hamiltonian, as has been
illustrated in Appendix~\ref{app:zero}.

\section{Renormalization group theory of critical properties}
\label{sec:rg}

There is already a well-established theory\cite{bz} for the bulk
phase transition at $g_0=g_c$. Here we will show how this theory
can be extended to the impurity correlations. This will be done
with a single additional impurity wavefunction renormalization
constant $Z^{\prime}$---from the perspective of boundary critical
phenomena, this is a boundary renormalization factor at the
impurity site $x=0$. As noted earlier, the Berry phase in
$\mathcal{S}_{\text{imp}}$ is an exactly marginal perturbation to
the bulk critical point: it is protected by its topological
nature, and hence there is no additional coupling constant
renormalization associated with the impurity spin. We will use
this critical theory to obtain results for the impurity and local
susceptibilities at $T>0$, and for the field dependence of the
free energy at $T=0$.

First, let us recall the bulk renormalization theory from
Ref.~\onlinecite{bz}. There is a field renormalization factor,
$Z$, defined by
\begin{equation}
N_{\alpha} (x, \tau) = \sqrt{Z} N_{R,\alpha} (x, \tau)~~~;~~~x\neq
0, \label{r1}
\end{equation}
where $N_{R \alpha}$ is the renormalized field. In the present
quantum impurity context, this renormalization will be adequate at
all spatial points away from the impurity, as has been indicated
above. Second, Ref.~\onlinecite{bz} has a coupling constant
renormalization
\begin{equation}
g_0 = \frac{g Z_1 \mu^{1-d}}{S_{d+1}} \label{r2}
\end{equation}
where $g$ is the renormalized dimensionless coupling constant,
$\mu$ is a cut-off momentum scale, and
\begin{equation}
S_d \equiv \frac{2 \pi^{d/2}}{(2 \pi)^d \Gamma (d/2)} \label{r3}
\end{equation}
is a phase-space factor. To two-loop order and in the minimal
subtraction scheme, these bulk renormalization constants are given
by \cite{bz}
\begin{eqnarray}
Z &=& 1 + \frac{2g}{\epsilon} + \frac{3 g^2}{\epsilon^2} \nonumber
\\
Z_1 &=& 1 + \frac{g}{\epsilon} +
\frac{g^2}{\epsilon^2}(1+\epsilon/2) \label{r4}
\end{eqnarray}
where
\begin{equation}
\epsilon = d-1. \label{epsilon}
\end{equation}
These
constants give the beta function
\begin{equation}
\beta(g) = \epsilon g - g^2 - g^3 \label{r5}
\end{equation}
which has a fixed point at $g=g^{\ast}$ which describes the bulk
quantum critical point, with
\begin{equation}
g^{\ast} = \epsilon - \epsilon^2 + \mathcal{O}(\epsilon^3)
\label{r6}
\end{equation}

Let us now turn to the impurity correlations. These require only
an additional boundary wavefunction renormalization which we
define by
\begin{equation}
N_{\alpha} (x=0, \tau) = \sqrt{Z^{\prime}} N_{R,\alpha} (x=0,
\tau). \label{r7}
\end{equation}
We discuss the computation of $Z^{\prime}$ in
Appendix~\ref{app:eta}, where we find
\begin{equation}
\frac{Z^{\prime}}{Z} = 1 - \frac{2 \pi^2 g^3 S^2}{3 \epsilon} +
\mathcal{O}(g^4) \label{r8}
\end{equation}
This renormalization constant implies that impurity spin
correlations behave as
\begin{equation}
\left\langle N_{\alpha} (x=0,\tau) N_{\alpha} (x=0,0)
\right\rangle \sim \frac{1}{\tau^{\eta^{\prime}}}~~~~,~~g_0=g_c
\label{r9}
\end{equation}
where
\begin{eqnarray}
\eta^{\prime} &=& \epsilon+\eta+\left. \beta(g) \frac{d \ln
Z^{\prime}}{dg} \right|_{g=g^{\ast}} \nonumber \\
&=& \epsilon+\eta - 2 \pi^2 S^2 \epsilon^3
+\mathcal{O}(\epsilon^4). \label{r10}
\end{eqnarray}
Here $\eta$ is the nearly-vanishing anomalous dimension of the
bulk critical point which was mentioned in
Section~\ref{sec:ft}---it controls the decay of $N_{\alpha}$
correlations sufficiently far away from the impurity:
\begin{equation}
\left\langle N_{\alpha} (x,\tau) N_{\alpha} (x,0) \right\rangle
\sim \frac{1}{\tau^{\epsilon+\eta}}~~~~,~~g_0=g_c~~,~~x
\rightarrow \infty \label{r11}
\end{equation}

The results in Appendix~\ref{app:impt} can also be easily used to
obtain an $\epsilon$ expansion for the universal constant
$\mathcal{C}_1$ in (\ref{e5}) which determines the anomalous Curie
response of $\chi_{\text{imp}}$ at the critical point. We begin by
substituting the renormalized coupling $g$ defined by (\ref{r2})
into the dimensionally regularized expression for
$\chi_{\text{imp}}$ defined by (\ref{a8}), (\ref{mexp}),
(\ref{perpexp}), and (\ref{parexp}). We expand the resulting
expression to order $g^2$, and then expand the coefficient of each
such term in powers of $\epsilon$. Consistency of the theory
demands that poles in $\epsilon$ cancel at this point, and this is
indeed the case. Finally we substitute the fixed-point value
$g=g^{\ast}$ in (\ref{r6}) into this expression. We find that all
dependence upon $\mu$ disappears at this point, which is strong
evidence for the universality expressed by (\ref{e2}); the final
expression then yields
\begin{equation}
\mathcal{C}_1 = \frac{S^2}{3} \left( 1 + 2 \epsilon +
\left(1+\frac{\pi^2}{12}\right)\epsilon^2 +
\mathcal{O}(\epsilon^3) \right). \label{r12}
\end{equation}
As is also the case with the bulk exponents in the $d=1+\epsilon$
expansion, we do not expect the estimate of (\ref{r12}) to be
accurate at $\epsilon=1$. As discussed in I, we expect on physical
grounds that $S^2/3 < \mathcal{C}_1 < S(S+1)/3$.

Next we consider the local susceptibility. As in I, this diverges
near the critical point as
\begin{equation}
\chi_{\text{loc}} \sim T^{-1+\eta^{\prime}} \Phi_{\text{loc}}
\left( \frac{T}{\rho_s^{1/(d-1)}} \right), \label{r13}
\end{equation}
with $\Phi_{\text{loc}}$ a universal scaling function; the small
argument behavior of $\Phi_{\text{loc}}$ should be compatible with
(\ref{loc2}), while its infinite argument limit is a constant. By
analysis similar to that outlined in the previous paragraph, the
expressions in Appendix~\ref{app:loc} can be verified to be
consistent with (\ref{r13}) and the value of $\eta^{\prime}$ in
(\ref{r10}).

Finally, we turn to the response to an applied field at $T=0$,
discussed earlier in Section~\ref{sec:zero} and also in
Appendix~\ref{app:zero}. At the critical point, any applied
$H_{\perp}$ will induce long-range magnetic order in the
bulk\cite{dsz} for $d>1$. The scaling form (\ref{x3}) nevertheless
holds as $\rho_s \rightarrow 0$, and we therefore obtain
\begin{equation}
\mathcal{F}_{\text{imp}} (H_{\perp}) - \mathcal{F}_{\text{imp}}
(0) = -\mathcal{C}_{\mathcal{F}} H_{\perp}~~~~;~~~~g_0 = g_c
\label{q1}
\end{equation}
where $\mathcal{C}_{\mathcal{F}} \equiv \Phi_{\mathcal{F}}
(\infty)$ is a universal number. This universal number can be
obtained directly from (\ref{zero2}) by the methods discussed
above for $\chi_{\text{imp}}$, and we obtain
\begin{equation}
\mathcal{C}_{\mathcal{F}} = \frac{\pi S^2}{2} \epsilon +
\mathcal{O} ( \epsilon^2 ) \label{q2}
\end{equation}

\section{Conclusions}
\label{sec:conc}

This paper has introduced the field theory (\ref{simpn}) as a
description of the low temperature properties of arbitrary static
impurities in quantum antiferromagnets. The bulk fluctuations of
the antiferromagnet are described by the familiar O(3) quantum
non-linear sigma model. Remarkably, this venerable and strongly
interacting field theory permits an exactly marginal perturbation,
albeit on a `boundary', which has not been noticed before: this is
the topological Berry phase of a spin $S$ impurity. We have
computed here the physical consequences of this marginal
perturbation and so obtained a new description of the spin
dynamics of the impurity.

A preliminary comparison has been made\cite{anders} between  our
theoretical result (\ref{e4}) and the numerical results of
Ref.~\onlinecite{sandvik}. Reasonable agreement is found for the
impurity susceptibility of a vacancy, and a more detailed
comparison will appear later.

Near the quantum critical point associated with the loss of
long-range antiferromagnetic order, our results were obtained in
an expansion in $(d-1)$. Key scaling features of these results
were found to be in good accord with those obtained earlier in a
$(3-d)$ expansion in I. In particular, right at the critical
point, we confirmed the existence of a Curie $1/T$ impurity spin
susceptibility but with an anomalous Curie constant not given by
an integer or half-odd-integer spin. Our new result for the Curie
constant is in (\ref{r12}). Although the numerical estimates for
critical properties obtained from the $(d-1)$ expansion seem
rather unreliable, this expansion nevertheless provides convincing
evidence for the existence of a strongly-coupled impurity fixed
point with the physical properties discussed herein and in I.

\begin{acknowledgments}
We are very grateful to Anders Sandvik and Kaj H\"oglund for
informing us about their results prior to publication and for
useful discussions. We also thank Oleg Sushkov for valuable
discussions; he has obtained results \cite{sushkov2} which overlap
with the present paper. This research was supported by US NSF
Grant DMR 0098226 and the DFG Center for Functional Nanostructures
at the University of Karlsruhe.
\end{acknowledgments}

\appendix

\section{Diagrammatic perturbation theory}
\label{app:perturb}

We will consider perturbation theory in the presence of an applied
uniform magnetic field $H_{\alpha}$ under which (\ref{simpn}) is
modified by
\begin{eqnarray}
\partial_{\tau} N_{\alpha} &\rightarrow & \partial_{\tau}
N_{\alpha} - i \epsilon_{\alpha\beta\gamma}H_{\beta} N_{\gamma}
\nonumber \\
\mathcal{S}_{\text{imp}} & \rightarrow & \mathcal{S}_{\text{imp}}
- S H_{\alpha} \int_0^{1/T} d \tau N_{\alpha} (x=0,\tau)
\label{a1}
\end{eqnarray}
This construction ensures that $H_{\alpha}$ couples to a conserved
total spin of the Hamiltonian.

\begin{figure}
\centerline{\includegraphics[width=2.5in]{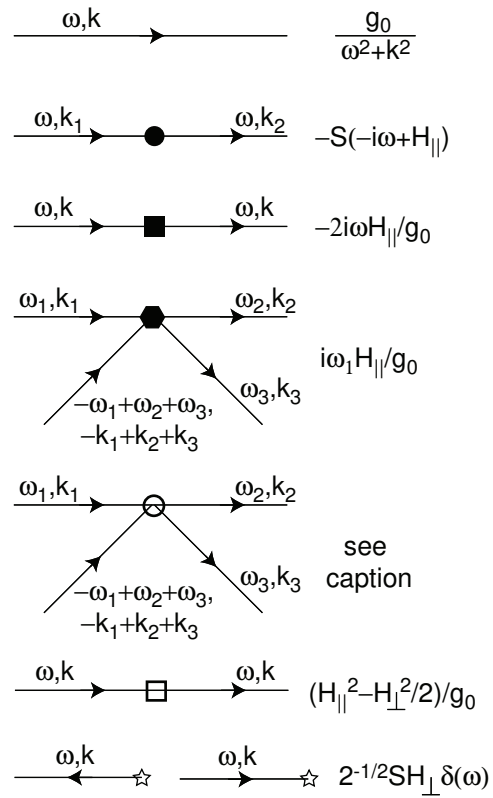}}
\caption{Propagator and vertices appearing in the calculation to
the order needed. The weight of the fifth term is $(\vec{k}_2
\cdot \vec{k}_3 + \omega_2 \omega_3 + \vec{k}_1 \cdot ( \vec{k}_2
+ \vec{k_3} - \vec{k}_1) + \omega_1 ( \omega_2 + \omega_3 -
\omega_1))/(4 g_0)$.} \label{figfeynman}
\end{figure}

As in Refs.~\onlinecite{bz,book}, the perturbation theory in $g_0$
is generated by assuming that $N_{\alpha}$ is locally polarized
along a particular direction (say (0,0,1)), and by expanding in
deviations of $N_{\alpha}$ about this direction. We do this here
with the following parametrization in terms of a complex field
$\psi$, adapted from the Holstein-Primakoff representation:
\begin{equation}
N_{\alpha} = \left( \frac{\psi+\psi^{\ast}}{2} \sqrt{2 - |\psi
|^2}, \frac{\psi- \psi^{\ast}}{2 i} \sqrt{2 - |\psi|^2}, 1 -
|\psi|^2 \right). \label{a2}
\end{equation}
The advantage of the representation (\ref{a2}) is that with the
gauge choice
\begin{equation}
A_{\alpha} (n) = \frac{1}{1 + n_z}\left(- n_y, n_x, 0 \right),
\label{a3}
\end{equation}
the Berry phase takes the following simple exact form
\begin{equation}
i A_{\alpha} (n) \frac{d n_{\alpha}}{d \tau} = \frac{1}{2} \left(
\psi^{\ast} \frac{\partial\psi}{\partial \tau}- \psi
\frac{\partial\psi^{\ast}}{\partial \tau}\right), \label{a4}
\end{equation}
where the right-hand-side is to be evaluated at $x=0$.
Furthermore, the measure term in the functional integral also has
the simple form
\begin{equation}
\int \mathcal{D} N_{\alpha} \delta\left( N_{\alpha}^2 - 1 \right)
= \frac{1}{2} \int \mathcal{D} \psi \mathcal{D} \psi^{\ast}
\label{a5}
\end{equation}
The remaining terms in the action are obtained by inserting
(\ref{a1},\ref{a2}) into (\ref{simpn}) and expanding the results
in powers of $\psi$---this yields a number of non-linearities
which are analogous to those that appear in Ref.~\onlinecite{bz},
and these can be used to generate a Feynman graph expansion in a
similar manner. We summarize the propagator and the vertices, to
the order needed in our computation here, in
Fig.~\ref{figfeynman}.

\begin{figure}
\centerline{\includegraphics[width=1.1in]{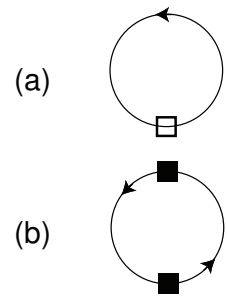}}
\caption{Diagrams for the bulk susceptibilities to order $g_0^0$.}
\label{figbulk}
\end{figure}

First, we recall the results for the bulk response, in the absence
of the impurity. The free energy is expanded as in (\ref{a7}), and
this leads to the diagrams in Fig~\ref{figbulk} to order $g_0^0$.
There is no bulk linear dependence on $H_{\alpha}$ to all orders
in $g_0$, and hence $m=0$ in the absence of the impurity. To
quadratic order in $H_{\alpha}$ we have the bulk susceptibilities
(per unit volume)
\begin{eqnarray}
\chi_{\perp,\text{b}} &=& \frac{1}{g_0} + (\ref{figbulk}a) \nonumber \\
(\ref{figbulk}a) &=& - T \sum_{\omega_n \neq 0} \int \frac{d^d
k}{(2 \pi)^d} \frac{1}{\omega_n^2 + k^2} \nonumber \\
&=& - 2T^{d-1} \mathcal{J}_1 (2 \pi)^{d-2} \zeta (2-d)
\label{perpb}
\end{eqnarray}
and
\begin{eqnarray}
\chi_{\parallel,\text{b}} &=&  (\ref{figbulk}a)+ (\ref{figbulk}b) \nonumber \\
(\ref{figbulk}a) &=& 2 T \sum_{\omega_n \neq 0} \int \frac{d^d
k}{(2 \pi)^d} \frac{1}{\omega_n^2 + k^2} \nonumber \\
&=& 4 T^{d-1} \mathcal{J}_1 (2 \pi)^{d-2} \zeta (2-d) \nonumber \\
(\ref{figbulk}b) &=& -4 T \sum_{\omega_n \neq 0} \int \frac{d^d
k}{(2
\pi)^d} \frac{\omega_n^2}{(\omega_n^2 + k^2)^2} \nonumber \\
&=& -8 T^{d-1} \mathcal{J}_2 (2 \pi)^{d-2} \zeta (2-d),
\label{parb}
\end{eqnarray}
where
\begin{equation}
\mathcal{J}_a \equiv  \frac{\Gamma(a-d/2)}{(4 \pi)^{d/2} \Gamma
(a) }.
\end{equation}

As discussed in Section~\ref{sec:perturb}, all intermediate
Matsubara frequencies in all diagrams in this appendix are summed
only over non-zero values; the integration over the zero Matsubara
frequency modes leads to (\ref{a8}). There are no infrared
divergences in any graph (because of the summation over non-zero
Matsubara frequencies), while ultraviolet divergences appear in
individual graphs for $d \geq 1$. We also list the expressions for
the individual graphs obtained in the dimensional regularization
method, obtained by analytic continuation from the $d<1$
region---these will be useful in our renormalization group
analysis. The dimensionally-regularized expressions ware obtained
by first performing the momentum integrations, and the frequency
summations are then naturally expressed in terms of the Riemann
zeta function $\zeta (s)$. There are also many sensitive
cancellations in the ultraviolet divergences of the various graphs
considered in this appendix, and these will appear as cancellation
of poles in the dimensionally regularized expressions. In
Section~\ref{sec:perturb} we have also considered the expressions
of this appendix directly in $d=2$ without dimensional
regularization, and these results illustrate the cancellation of
ultraviolet divergences upon expression of the results in terms of
physical observables.

The application of the perturbation theory towards computation of
physical properties of the impurity in different regimes will be
presented in separate subsections below.

\subsection{Impurity susceptibility at $T>0$}
\label{app:impt}

We first address the computation of the impurity magnetic
susceptibility, $\chi_{\text{imp}}$ at non-zero temperatures.

\begin{figure}
\centerline{\includegraphics[width=1.6in]{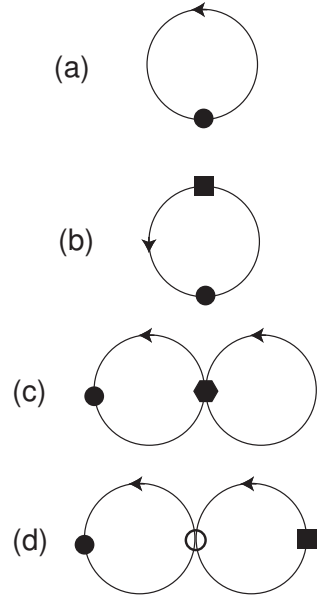}}
\caption{Diagrams contributing to $m$ to order $g_0^2$.}
\label{figm}
\end{figure}

\begin{figure}
\centerline{\includegraphics[width=0.7in]{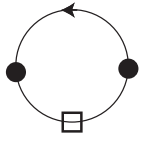}}
\caption{Diagram contributing to $\chi_{\perp,\text{imp}}$ to
order $g_0^2$ at $T>0$.} \label{figperp}
\end{figure}

\begin{figure}
\centerline{\includegraphics[width=1.1in]{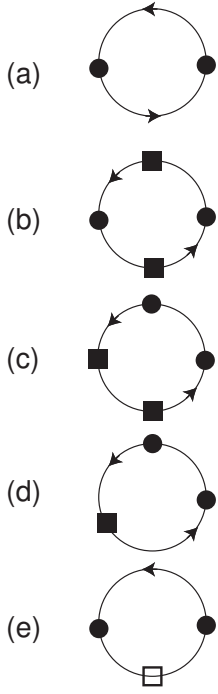}}
\caption{Diagrams contributing to $\chi_{\parallel,\text{imp}}$ to
order $g_0^2$.} \label{figpar}
\end{figure}

The diagrams for the perturbative expressions for the impurity
contributions to the quantities in (\ref{a7}) are shown in
Figs~\ref{figm}-\ref{figpar}.
Now there is a contribution to linear order in $H_{\parallel}$,
and Fig~\ref{figm} yields the following expressions for $m$:
\begin{eqnarray}
m &=& S + (\ref{figm}a) + (\ref{figm}b) + (\ref{figm}c) + (\ref{figm}d) \nonumber \\
(\ref{figm}a) &=& -S g_0 T \sum_{\omega_n \neq 0} \int \frac{d^d
k}{(2
\pi)^d} \frac{1}{\omega_n^2 + k^2} \nonumber \\
&=& - 2 S g_0 T^{d-1} \mathcal{J}_1 (2 \pi)^{d-2} \zeta (2-d) \nonumber \\
(\ref{figm}b) &=& 2 S g_0 T \sum_{\omega_n \neq 0} \int \frac{d^d
k}{(2
\pi)^d} \frac{\omega_n^2}{(\omega_n^2 + k^2)^2} \nonumber \\
&=& 4 S g_0 T^{d-1} \mathcal{J}_2 (2 \pi)^{d-2} \zeta (2-d) \nonumber \\
(\ref{figm}c) &=& -2 S g_0^2 T^2 \left[\sum_{\omega_n \neq 0} \int
\frac{d^d
k_1}{(2 \pi)^d}  \frac{\omega_n^2}{(\omega_n^2 + k_1^2)^2} \right]  \nonumber \\
&~&~~~~~~~~~~~~\times \left[\sum_{\epsilon_n \neq 0} \int
\frac{d^d k_2}{(2 \pi)^d}  \frac{1}{\epsilon_n^2 + k_2^2} \right]
\nonumber \\
 &=& - 8 S g_0^2 T^{2d-2} \mathcal{J}_1
\mathcal{J}_2 \left[ (2 \pi)^{d-2} \zeta (2-d) \right]^2 \nonumber \\
 (\ref{figm}d) &=&  4 S g_0^2 T^2 \left[\sum_{\omega_n \neq 0} \int
\frac{d^d k}{(2 \pi)^d} \frac{\omega_n^2}{(\omega_n^2 + k^2)^2}
\right]^2 \nonumber \\
&=& 16 S g_0^2 T^{2d-2} \mathcal{J}_2^2 \left[ (2 \pi)^{d-2} \zeta
(2-d) \right]^2 . \label{mexp}
\end{eqnarray}

Similarly, for $\chi_{\perp,\text{imp}}$ we only have the diagram
in Fig~\ref{figperp} which yields:
\begin{eqnarray}
\chi_{\perp,\text{imp}} &=& (\ref{figperp}) \nonumber \\
(\ref{figperp}) &=&  S^2 g_0^2 T \sum_{\omega_n \neq 0} \int
\frac{d^d k_1}{(2 \pi)^d} \frac{d^d k_2}{(2 \pi)^d}
\frac{\omega_n^2}{(\omega_n^2 +
k_1^2)^2} \nonumber \\
&~&~~~~~~~~~~~~~~\times\frac{1}{ (\omega_n^2 + k_2^2)} \nonumber
\\
&=& 2 S^2 g_0^2 T^{2d-3} \mathcal{J}_1 \mathcal{J}_2 (2
\pi)^{2d-4} \zeta(4-2d). \label{perpexp}
\end{eqnarray}

Finally, for $\chi_{\parallel,\text{imp}}$, we have the diagrams
in Fig~\ref{figpar}, from which we obtain:
\begin{eqnarray}
\chi_{\parallel,\text{imp}} &=& (\ref{figpar}a) + (\ref{figpar}b) + (\ref{figpar}c) + (\ref{figpar}d) +
(\ref{figpar}e) \nonumber \\
(\ref{figpar}a) &=&  S^2 g_0^2 T \sum_{\omega_n \neq 0} \int
\frac{d^d k_1}{(2 \pi)^d} \frac{d^d k_2}{(2 \pi)^d}
\frac{1}{(\omega_n^2 +
k_1^2)} \nonumber \\
&~&~~~~~~~~~~~~~~\times\frac{1}{ (\omega_n^2 + k_2^2)} \nonumber
\\
&=& 2 S^2 g_0^2 T^{2d-3} \mathcal{J}_1^2 (2
\pi)^{2d-4} \zeta(4-2d) \nonumber \\
 (\ref{figpar}b) &=&  4 S^2
g_0^2 T \sum_{\omega_n \neq 0} \int \frac{d^d k_1}{(2 \pi)^d}
\frac{d^d k_2}{(2 \pi)^d} \frac{\omega_n^4}{(\omega_n^2 +
k_1^2)^2} \nonumber \\
&~&~~~~~~~~~~~~~~\times\frac{1}{ (\omega_n^2 + k_2^2)^2} \nonumber
\\
&=& 8 S^2 g_0^2 T^{2d-3}  \mathcal{J}_2^2 (2
\pi)^{2d-4} \zeta(4-2d) \nonumber \\
(\ref{figpar}c) &=& 8 S^2 g_0^2 T \sum_{\omega_n \neq 0} \int
\frac{d^d k_1}{(2 \pi)^d} \frac{d^d k_2}{(2 \pi)^d}
\frac{\omega_n^4}{(\omega_n^2 +
k_1^2)^3} \nonumber \\
&~&~~~~~~~~~~~~~~\times\frac{1}{ (\omega_n^2 + k_2^2)} \nonumber
\\
&=& 16 S^2 g_0^2 T^{2d-3} \mathcal{J}_3 \mathcal{J}_1 (2
\pi)^{2d-4} \zeta(4-2d) \nonumber \\
(\ref{figpar}d) &=& - 8 S^2 g_0^2 T \sum_{\omega_n \neq 0} \int
\frac{d^d k_1}{(2 \pi)^d} \frac{d^d k_2}{(2 \pi)^d}
\frac{\omega_n^2}{(\omega_n^2 +
k_1^2)^2} \nonumber \\
&~&~~~~~~~~~~~~~~\times\frac{1}{ (\omega_n^2 + k_2^2)} \nonumber
\\
&=& - 16 S^2 g_0^2 T^{2d-3} \mathcal{J}_1 \mathcal{J}_2 (2
\pi)^{2d-4} \zeta(4-2d) \nonumber \\
(\ref{figpar}e) &=& - 2 S^2 g_0^2 T \sum_{\omega_n \neq 0} \int
\frac{d^d k_1}{(2 \pi)^d} \frac{d^d k_2}{(2 \pi)^d}
\frac{\omega_n^2}{(\omega_n^2 +
k_1^2)^2} \nonumber \\
&~&~~~~~~~~~~~~~~\times\frac{1}{ (\omega_n^2 + k_2^2)} \nonumber
\\
&=& -4 S^2 g_0^2 T^{2d-3} \mathcal{J}_1 \mathcal{J}_2 (2
\pi)^{2d-4} \zeta(4-2d) \label{parexp}
\end{eqnarray}

\subsection{Local susceptibility at $T>0$}
\label{app:loc}

The response to a field applied only near the impurity site can be
computed as in Appendix~\ref{app:loc}. Only the graphs in
Fig~\ref{figm}a and ~\ref{figpar}a now contribute, and we have
therefore
\begin{eqnarray}
m_{\text{loc}} &=& S + (\ref{figm}a) \nonumber \\
\chi_{\parallel,\text{loc}} &=& (\ref{figpar}a) \nonumber \\
\chi_{\perp,\text{loc}} &=& 0 \label{apploc1}
\end{eqnarray}
where the values of the respective graphs are as specified in
(\ref{mexp}) and (\ref{parexp}).

\subsection{Spin correlations at $T=0$}
\label{app:eta}

The methods above can also be extended to obtain impurity spin
correlations at $T=0$ and $g_0=g_c$. As long as we restrict
ourselves to rotationally invariant correlation functions, direct
perturbation theory in $g_0$ is free of infrared divergences.
\begin{figure}
\centerline{\includegraphics[width=2.1in]{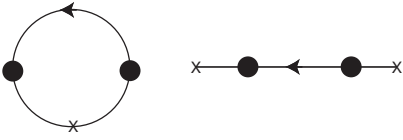}}
\caption{Lowest order diagram contributing to the correlator in
(\protect\ref{a11}) which depends upon the presence of the
impurity. The $X$'s denote external sources for the fields.}
\label{figeta}
\end{figure}
For the impurity spin correlation in (\ref{r9}), the first
corrections which depend upon the presence of the impurity do not
appear until order $g_0^3$: these arise from the graphs shown in
Fig~\ref{figeta} and lead to the following expression:
\begin{eqnarray}
&&\left\langle N_{\alpha} (x=0,\tau) N_{\alpha} (x=0,0)
\right\rangle = 1 + \nonumber \\
&&~~2g_0^3 S^2 \int \frac{d \omega}{2 \pi} \frac{d^d k_1}{(2
\pi)^d}\frac{d^d k_2}{(2 \pi)^d} \frac{d^d k_3}{(2 \pi)^d}
\nonumber \\
&&~~~~\times \frac{\omega^2 (1 - \cos (\omega \tau))}{(\omega^2 +
k_1^2)(\omega^2 + k_3^2)(\omega^2 + k_3^2)}+ \ldots. \label{a11}
\end{eqnarray}
Here the ellipses denote numerous lower-order terms which do not
depend upon the presence of the impurity and hence are the same at
$x=0$ and $x \neq 0$; the first term which breaks translational
invariance is shown in (\ref{a11}). The integrals in (\ref{a11})
can be easily evaluated in dimensional regularization, and the
second term in (\ref{a11}) equals
\begin{equation}
- \frac{2 g_0^3 S^2 [\Gamma((2-d)/2)]^3 \Gamma(3d-3)
\cos(3\pi(d-1)/2)}{(4 \pi)^{3d/2} \pi \tau^{3d-3}} \label{a12}
\end{equation}
Picking out the pole in $\epsilon$ in (\ref{a12}), we immediately
obtain (\ref{r8}).

\subsection{Response to a field at $T=0$}
\label{app:zero}

As discussed in Section~\ref{sec:zero}, we need the impurity
contribution to the free energy in the presence of an applied
transverse magnetic field $H_{\perp}$, $\mathcal{F}_{\text{imp}}
(H_{\perp})$. We will see that the response is singular as
$H_{\perp} \rightarrow 0$. The singularity can be cutoff by an
easy-axis spin anisotropy, and for completeness, we perform the
computation in the presence of such an anisotropy. So we modify
the action by
\begin{equation}
\mathcal{S}_b [N_{\alpha}] \rightarrow \mathcal{S}_b [N_{\alpha}]
-\frac{D}{2} \int d^d x \int_0^{1/T} d\tau N_z^2 . \label{zero1}
\end{equation}

\begin{figure}
\centerline{\includegraphics[width=1.3in]{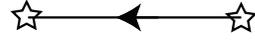}}
\caption{Diagram contributing to the free energy
$\mathcal{F}(H_{\perp})$ to order $g_0$. at $T=0$. The propagator
represents the $x$ component of $N_{\alpha}$ and includes the open
square vertex in Fig~\protect\ref{figfeynman} to all orders in
$H_{\perp}$, along with the easy-axis anisotropy in
(\protect\ref{zero1}); it equals $g_0/(\omega^2 + k^2 + D+
H_{\perp}^2)$.} \label{figzero}
\end{figure}

Because we are now computing the free energy to all orders in the
applied field, the Feynman graph expansion is quite tedious, and
we will be satisfied by obtaining the result only to order $g_0$.
The computation is done most simply using the Cartesian components
$\sim (\psi + \psi^{\ast}, -i(\psi - \psi^{\ast}))$, and to
leading order in $g_0$, only the graph shown in Fig~\ref{figzero}
contributes.
Note that this diagram did not appear in the computation at $T>0$
because of the restriction there to summation over non-zero
Matsubara frequencies and the delta function in frequency
associated with the vertex in Fig~\ref{figzero}; the leading term
in $\chi_{\perp,\text{imp}}$ was of order $g_0^2$ at $T>0$. From
the diagram in Fig~\ref{figzero} we obtain
\begin{eqnarray}
\mathcal{F}_{\text{imp}} (H_{\perp}) &=& -\frac{g_0 H_{\perp}^2
S^2}{2} \int \frac{d^d
k}{(2 \pi)^d} \frac{1}{k^2 + D + H_{\perp}^2} \nonumber \\
&=& -\frac{g_0 H_{\perp}^2 S^2}{2} \mathcal{J}_1 (D +
H_{\perp}^2)^{(d-2)/2}. \label{zero2}
\end{eqnarray}
This graph has a log singularity in $d=2$. The same logarithm
appeared in a different manner in the $T>0$ computation: it was
present in Fig~\ref{figm}a. Ultimately it is only
$\chi_{\text{imp}}$ that is physically measurable at $T>0$, and it
is clear now that the logarithm appears in different places
depending upon the different organizations of perturbation theory
at $T=0$ and $T>0$.

\section{Perturbation theory for general $\gamma$}
\label{app:gamma}

The computations elsewhere in this paper have been limited to the
case in which the coupling between the impurity spin and the bulk
antiferromagnetic spin fluctuations, $\gamma$, has effectively
been sent to infinity. We have argued that this limit appears
naturally in the vicinity of the quantum critical point. This
appendix will consider the general $\gamma$ case, and consider the
extent to which the low $T$ properties away from the critical
point are independent of the value of $\gamma$.

We shall be concerned here with the partition function
\begin{eqnarray}
&& \mathcal{Z}_{\gamma} = \int \mathcal{ D} N_{\alpha} (x, \tau)
\delta \left( N_{\alpha}^2 - 1 \right) \mathcal{D}n_{\alpha}
\delta( n_{\alpha}^2 - 1) \nonumber \\
&&~~~~~~~~~~~~~~\times\exp\left(- \mathcal{S}_{\text{b}}
[N_{\alpha}] - \mathcal{S}_{\text{imp},\gamma} \right) \nonumber \\
&& \mathcal{S}_{\text{imp},\gamma} =  \int_0^{1/T} d \tau \bigg[ i
S
A_{\alpha} (n) \frac{d n_{\alpha}(\tau)}{d \tau} \nonumber \\
&&~~~~~~~~~~~~~~~~~~~ - \gamma S N_{\alpha} (x = 0, \tau)
n_{\alpha} (\tau) \bigg], \label{actiongamma}
\end{eqnarray}
where $\mathcal{S}_{\text{b}}[N_{\alpha}]$ is as in (\ref{simpn}),
and $A_{\alpha} (n)$ is defined by (\ref{diracmono}). In
principle, it is possible to generate an expansion in powers of
$g_0$, with each term containing its exact dependence on $\gamma$
and $S$; this requires an exact treatment of the impurity spin
fluctuations, and this can be done by the method described in
Appendix C of I. Here, we shall use the method described above in
Section~\ref{app:perturb} with a parametrization similar to
(\ref{a2}) applied also to $n_{\alpha}$. By this method, it is not
difficult to obtain results order-by-order in $g_0$, dropping only
diagrams with a `tadpole' factor of the impurity spin propagator
({\em i.e.} with a simple closed loop of the impurity spin
propagator)---these are easily seen to have a prefactor of
$e^{-\gamma/T}$ (we assume, without loss of generality, that
$\gamma > 0$).

We now present results to order $g_0$ for the impurity spin
susceptibility at $T>0$, computed above in
Appendix~\ref{app:impt}. We will omit all details and merely
present final results to leading order in $g_0$. Dropping terms
with a pre-factor of $e^{-\gamma/T}$, we found
\begin{eqnarray}
&&m = S - \gamma^2 S g_0 T \sum_{\omega_n \neq 0} \int \frac{d^d
k}{(2 \pi)^d}\Biggl[ \nonumber \\
&& \frac{1}{(\omega_n^2 + k^2)(-i \omega_n + \gamma)^2} + \frac{2
i \omega_n}{(\omega_n^2 + k^2)^2 (-i
\omega_n+ \gamma)} \Biggr] \nonumber \\
&& \chi_{\parallel,\text{imp}} =  2\gamma S g_0 T \sum_{\omega_n
\neq 0} \int \frac{d^d k}{(2 \pi)^d}\Biggl[ \nonumber
\\
&& \frac{4 \omega_n^2}{(\omega_n^2+k^2)^3} -
\frac{1}{(\omega_n^2+k^2)^2} + \frac{\gamma}{(\omega_n^2 + k^2)
(-i \omega_n + \gamma)^3} \nonumber \\ && +
\frac{\gamma}{(\omega_n^2 + k^2)^2 (-i \omega_n + \gamma)} +
\frac{2 i\gamma \omega_n}{(\omega_n^2 + k^2)^2 (-i \omega_n +
\gamma)^2} \nonumber
\\ &&- \frac{4 \gamma \omega_n^2}{(\omega_n^2 + k^2)^3 (-i
\omega_n +
\gamma)}\Biggr] \nonumber \\
&& \chi_{\perp,\text{imp}} = \gamma S g_0 T \sum_{\omega_n
\neq 0} \int \frac{d^d k}{(2 \pi)^d} \Biggl[ \frac{1}{(\omega_n^2 + k^2)^2}\nonumber \\
&&~~~~~~ - \frac{\gamma}{(\omega_n^2 + k^2)^2 (-i \omega_n +
\gamma)} \Biggr]. \label{gamma1}
\end{eqnarray}
It is now easy to check that the $\gamma \rightarrow \infty$ limit
of these expressions is finite, and indeed agrees precisely with
the order $g_0$ results for $\chi_{\text{imp}}$ obtained in
Section~\ref{sec:imp} and Appendix~\ref{app:impt}; this is a
non-trivial check of our computations. Evaluation of the frequency
summations in (\ref{gamma1}) is a tedious but straightforward
exercise. After this, we combine the results using (\ref{a8}), and
evaluate the momentum integrals at low $T$ as in
Section~\ref{sec:perturb}, while keeping $\gamma$ finite; in the
limit of $T \ll \gamma,\Lambda$ we obtain in $d=2$
\begin{eqnarray}
\chi_{\rm imp} &=& \frac{S^2}{3T} \Biggl[ 1  + \left(\frac{T}{ \pi
\rho_s} - \frac{T^2}{\pi S \rho_s \gamma} \right)\ln \left(
\frac{\Lambda}{T} \right) \Biggr]. \label{gamma2}
\end{eqnarray}
Notice that the co-efficient of the $(1/\rho_s) \ln(1/T)$ is
independent of $\gamma$, and that it agrees with (\ref{e4}). Also,
at finite $\gamma$, the $T \ln(1/T)$ term does acquire a
non-universal $\gamma$-dependent correction.

\section{Low temperature properties in $d=3$}
\label{app:d3}

This appendix briefly describes the extension of our results to
$d=3$. The bulk quantum critical point in $d=3$ does not satisfy
strong scaling properties, and so we will not consider it here. We
will focus only on the low $T$ properties within the magnetically
ordered state, well away from any quantum critical point.

Magnetic long-range order is present for a finite range of $T>0$,
and so the magnetic response remains anisotropic as $T \rightarrow
0$. The quantities $m$, $\chi_{\perp,\text{imp}}$,
$\chi_{\parallel,\text{imp}}$ retain their separate physical
identities, and can be measured separately.

The low $T$ expansions for $m$, $\chi_{\perp,\text{imp}}$,
$\chi_{\parallel,\text{imp}}$ are obtained as in
Appendix~\ref{app:perturb}. Indeed, now we need not separate the
$\omega_n =0$ and the $\omega_n \neq 0$ modes as there is
long-range order for $T>0$: the expressions in
Appendix~\ref{app:perturb} can therefore be used here, after
converting all frequency summations to run over both zero and
non-zero values of Matsubara frequencies. In this manner,
(\ref{be1}) is modified to
\begin{eqnarray}
m &=& S \left[ 1 - \frac{T}{\rho_s} \left\{\int \frac{d^3 k}{(2
\pi)^3} \frac{1}{4 T^2 \sinh^2 (k/(2T))} \right\}\right. \nonumber
\\ &~&~~~~\times \left\{1 + \frac{T}{\rho_s} \int \frac{d^3 p}{(2
\pi)^3} \left(\frac{1}{pT
(e^{p/T}-1)} \right.\right. \nonumber \\
&~&~~~~~~~~~~~~~~~~~~~\left.\left.\left.- \frac{1}{4 T^2 \sinh^2
(p/(2T))} \right)\right\}\right]. \label{d31}
\end{eqnarray}
Evaluating the momentum integrations, and re-inserting factors of
$c$, we obtain
\begin{equation}
m = S \left( 1 - \frac{T^2}{6 c \rho_s} + \frac{T^4}{72 c^2
\rho_s^2} + \ldots \right). \label{d32}
\end{equation}
Interestingly, the expression (\ref{d32}) can also be obtained
simply by setting $d=3$ in (\ref{be3}). For the finite $\gamma$
case, discussed in Appendix~\ref{app:gamma}, the $T^2$ term above
remains unchanged, while the $T^4$ term does acquire
$\gamma$-dependent corrections.

The results for $\chi_{\parallel,\text{imp}}$ and
$\chi_{\perp,\text{imp}}$ now follow from (\ref{parexp}) and
(\ref{perpexp}). Setting $d=3$ in the dimensionally regularized
expressions here, we find that the co-efficient of the universal
$T^3$ term vanishes for both quantities. However, there are
non-universal $T^2$ corrections for both
$\chi_{\parallel,\text{imp}}$ and $\chi_{\perp,\text{imp}}$, and
these have to be estimated directly from the expressions in
(\ref{parexp}) and (\ref{perpexp}): the frequency summations have
to be evaluated first (including the zero Matsubara frequencies),
and then the momentum integrations have to be evaluated with a
finite cutoff.

Similar techniques apply to the response to a local field
discussed in Appendix~\ref{app:loc}. Now we obtain the universal
correction
\begin{equation}
m_{\text{loc}} (T) - m_{\text{loc}} (0) = - \frac{ST^2}{12 c
\rho_s} \label{d33}
\end{equation}
along with non-universal $T^2$ corrections to
$\chi_{\perp,\text{loc}}$ and $\chi_{\parallel,\text{loc}}$. The
factor of 2 difference between the $T^2$ terms in (\ref{d33}) and
(\ref{d32}) is an interesting characteristic of the theory.

\section{Comment on Sushkov's computation}
\label{app:sushkov}

It has been claimed by Sushkov \cite{sushkov} that the Curie
constant remains $\mathcal{C}_1 = 1/4$ for a $S=1/2$ impurity at
the quantum critical point of an antiferromagnet. Here we show
using the model of his paper that there is a perturbative
correction to the impurity susceptibility, and that this implies
an anomalous Curie constant. Of course, the possibility remains
open that the $(3-d)$ and $(d-1)$ expansions both fail in $d=2$
near the critical point, but reasons for such a possible failure
do not appear in Sushkov's arguments.

Sushkov models the bulk spin fluctuations at the quantum critical
point using a $S=1$ boson $t_{\alpha}$, as in
Ref.~\onlinecite{bondops}. These bosons are coupled to the
external magnetic field $H$ (assumed oriented along the $z$ axis)
and to the impurity moment $\hat{S}_{\alpha}$. This gives us the
model considered by Sushkov:
\begin{eqnarray}
\mathcal{H} &=& \mathcal{H}_0 + \mathcal{H}_1 \nonumber \\
\mathcal{H}_0 &=& \sum_{{\bf k}} \sum_{m=0,\pm 1} ( \varepsilon
({\bf k}) - m H)
t^{\dagger}_{m} ({\bf k}) t_{m } ({\bf k}) - H \hat{S}_z \nonumber \\
\mathcal{H}_1 &=& \lambda {\phi}_{\alpha}\hat{S}_{\alpha}
\end{eqnarray}
where ${\bf k}$ is the momentum of the $t_{\alpha}$ bosons with
energy $\varepsilon ({\bf k})$, and
\begin{equation}
{\phi}_{\alpha} = \sum_{{\bf k}} \frac{1}{\sqrt{2 \varepsilon
({\bf k})}}( {t}_{\alpha} ({\bf k}) + {t}_{\alpha}^{\dagger} ({\bf
k}) )
\end{equation}
and
\begin{eqnarray}
t_{x} &=& (t_1 + t_{-1})/\sqrt{2} \nonumber \\
t_{y} &=& i(t_1 - t_{-1})/\sqrt{2} \nonumber \\
t_{z} &=& t_0
\end{eqnarray}
It can be checked that $H$ couples to the total spin, which
commutes with the Hamiltonian.

Now we compute the free energy, $\mathcal{F}$, in a power series
in $\lambda$ in arbitrary $H$. To second order in $\lambda$, this
is done by the familiar formula
\begin{eqnarray}
&& \mathcal{F} = -T \ln {\rm Tr}
e^{-\mathcal{H}_0/T}\nonumber \\
&&- T \int_0^{1/T} \!\! d \tau \int_0^{\tau} d \tau_1 \frac{{\rm
Tr} \left( e^{-\mathcal{H}_0/T} \mathcal{H}_1 (\tau) \mathcal{H}_1
(\tau_1) \right )}{{\rm Tr} e^{-\mathcal{H}_0/T}} \label{ae1}
\end{eqnarray}
where
\begin{equation}
\mathcal{H}_1 (\tau) = e^{\mathcal{H}_0 \tau} \mathcal{H}_1
e^{-\mathcal{H}_0 \tau} \label{ae2}
\end{equation}
Everything in (\ref{ae1}) and (\ref{ae2}) can be evaluated
analytically by simple means, and then we can perform the
integrals over $\tau$ and $\tau_1$ - this was done using the
computer program Mathematica for arbitrary $H$ and $S$, without
using any diagrammatic perturbation theory. Finally, we can expand
the result in powers of $H$ and obtain for the impurity
susceptibility
\begin{equation}
\chi_{\rm imp} = \frac{S(S+1)}{3T} + \frac{2\lambda^2
S(S+1)}{3T^2}  \sum_{{\bf k}} \frac{e^{\varepsilon ({\bf
k})/T}}{\varepsilon ({\bf k})^2 (e^{\varepsilon ({\bf k}) /T} -
1)^2}  \label{ae3}
\end{equation}
This result agrees precisely with that obtained using a
diagrammatic approach in I. It disagrees with that of Sushkov, who
did not obtain any correction to the first free moment term---he
does not appear to have considered the cross-correlation between
the bulk magnetization of the $t_{\alpha}$ and the impurity
magnetization. Note that this disagreement appears already at the
level of bare perturbation theory, and does not involve any of the
subtleties associated with approaching the scaling limit at the
critical point in the $\epsilon$ or $1/N$ expansions.

In the quantum disordered regime above the paramagnetic phase, we
can model $\varepsilon ({\bf k}) = \sqrt{c^2 {\bf k}^2 +
\Delta_T^2}$, where $\Delta_T \sim \Delta > 0 $ is the spin gap
\cite{vbs,bondops}; here (\ref{ae3}) predicts a contribution of
order $e^{-\Delta /T}$ to the susceptibility, and so the moment is
indeed precisely quantized at $S$.

However, the quantum critical region \cite{vbs} we have $\Delta_T
\sim T$, and then (\ref{ae3}) yields a contribution to the
susceptibility of order $\lambda^2 T^{d-4}$. For $d<3$, the
dimensionless combination $\lambda^2 T^{d-3}$ approaches a
universal value at the fixed point, and a universal irrational
correction to the Curie term applies, as shown in much detail in
I.


\end{document}